\documentclass[preprint,12pt]{aastex}

\def\hii{H\,{\sc ii}}
\def\hi{H\,{\sc i}} 
\def\kms{\relax \ifmmode {\,\rm km~s}^{-1}\else \,km~s$^{-1}$\fi}
\def\farcm{\hbox{$.\mkern-4mu^\prime$}}
\def\farcs{\hbox{$.\!\!^{\prime\prime}$}}

\def\arcmin{\hbox{$^\prime$}}

\def\secd#1.#2{ #1\farcs#2 }               

\newcommand {\ha}{H$\alpha$}
\newcommand {\hb}{H$\beta$}
\newcommand {\oi}{[O\,{\sc i}]}
\newcommand {\oiii}{[O\,{\sc iii}]}
\newcommand {\nii}{[N\,{\sc ii}]}
\newcommand {\sii}{[S\,{\sc ii}]}

\begin{document}

\title{A Critical Examination of Hypernova Remnant Candidates\\in M101. I. MF\,83}
\author{Shih-Ping Lai, You-Hua Chu\altaffilmark{1}, C.-H. Rosie Chen\altaffilmark{1}}
\affil{Astronomy Department, University of Illinois, 1002 W. Green Street,
Urbana, IL 61801; \\
slai@astro.uiuc.edu, chu@astro.uiuc.edu, c-chen@astro.uiuc.edu}
\altaffiltext{1}{Visiting astronomer, Kitt Peak National Observatory, 
National Optical Astronomy Observatories, operated by 
the Association of Universities for Research in Astronomy, Inc., 
under a cooperative agreement with the National Science Foundation.}
\author{Robin Ciardullo\altaffilmark{1}}
\affil{Department of Astronomy and Astrophysics, Pennsylvania State University,
525 Davey Laboratory, University Park, PA 16802; rbc@astro.psu.edu}
\and
\author{Eva K. Grebel\altaffilmark{2,3}}
\affil{Department of Astronomy, University of Washington, Seattle, WA 
98195-1580 \\ and \\ 
Max-Planck-Institut f\"ur Astronomie, K\"onigstuhl 17, D-69117 Heidelberg, 
Germany;\\ grebel@mpia-hd.mpg.de}
\altaffiltext{2}{Visiting astronomer, Michigan-Dartmouth-MIT Observatory.}
\altaffiltext{3}{Hubble Fellow.}

\newpage
\begin{abstract}
The SNR candidate MF\,83 in M101 is coincident with a very luminous 
X-ray source.  Based on the high X-ray luminosity, it has been 
suggested that MF\,83 is a ``hypernova remnant" requiring an 
explosion energy about two orders of magnitude higher than normal 
supernovae.  We have analyzed high-quality ground-based and $HST$ 
observations of MF\,83, and find that MF\,83 is a star formation 
region, consisting of a large ionized gas shell and four \hii\ 
regions along its periphery.  Continuum images show OB associations 
in these \hii\ regions and within the large shell.  The shell has
an expansion velocity of $\sim$50 \kms\ and a diameter of $\sim$270 pc.
The optical properties of this shell in MF\,83 are similar to those
of X-ray-bright superbubbles in the Large Magellanic Cloud.
If the X-ray emission is indeed diffuse, the implied thermal energy 
in MF\,83 is high, a few $\times$10$^{52}$ ergs.
This amount of thermal energy requires a large number of concentrated
supernova explosions or one powerful explosion.
Future X-ray observations with a high angular resolution are needed
to resolve the diffuse emission and point sources in MF\,83, in order 
to determine more accurately the thermal energy in the shell interior
and its required explosion energy.

\end{abstract}


\newpage

\section{Introduction}

It has been generally accepted that a supernova's explosion energy
carried by its ejecta to form a supernova remnant (SNR) is roughly 
10$^{51}$ ergs (Jones et al.\ 1998).  Upon analyzing
the X-ray emission from SNRs in the giant spiral galaxy M101, Wang (1999)
discovered two SNRs that appeared to require explosion energies one to two
orders of magnitude higher.  One of these SNRs, NGC\,5471B, is inside 
the giant \hii\ region NGC\,5471 (Skillman 1985; Chu \& Kennicutt 1986).
The other SNR, MF\,83 (notation from \markcite{MF97}Matonick \& Fesen 1997),
also appears to be associated with a star formation region 
(Chu, Chen, \& Lai 2000).

Similarly high energies have been frequently found in gamma-ray bursts 
(GRBs).  As some GRBs have afterglows several hundred times more luminous 
than the brightest supernovae, they have been dubbed ``hypernovae."  Some
GRBs have been observed to be near star-forming regions (Paczy\'nski 1998).
Because of the similarities in the energy requirement and the 
interstellar/stellar environment, Wang (1999) called the two 
super-energetic SNRs in M101 ``hypernova remnants,"  although the 
association of hypernova remnants to the GRBs has been recently 
questioned by Paczy\'nski (2000).

Regardless of the GRB connection, SNRs with such large explosion energies 
are unusual and therefore warrant further investigation.
To study the physical nature of the two hypernova remnants in M101, 
we have obtained deep images with ground-based telescopes, 
high-resolution images with the {\it Hubble Space Telescope}, 
and high-dispersion spectra with an echelle spectrograph.
In this paper (Paper I), we report on the hypernova remnant MF\,83. 
The hypernova remnant NGC\,5471B will be presented in Paper II.

The suggestion that MF\,83 is super-energetic is based mainly on its high
X-ray luminosity (Wang 1999).  However, none of the existing X-ray 
observations have resolved MF\,83.  It is
possible that point sources contribute significantly to MF\,83's X-ray 
emission, since massive X-ray binaries are frequently found in star 
formation regions.  Only optical observations can resolve MF\,83; 
therefore, we have used our new observations to examine the physical
structure of MF\,83 and to seek independent evidence for a powerful
explosion.  At optical wavelengths, MF\,83 is unusually large, 
with a diameter approaching $\sim$270 pc (Matonick \& Fesen 1997), if we 
adopt a distance of 7.2 Mpc for M101 (\markcite{St98}Stetson et al. 1998).  
No known SNRs in the Local Group are so large.  
It is therefore important to measure the expansion velocity of MF\,83
in order to assess the validity of its identification as a SNR.

In this paper we report on our investigation of the environment and
energetics of the hypernova remnant MF\,83 using optical observations.  
The new observations are described in \S 2, and the analysis and results 
are presented in \S 3.  We critically examine the physical properties and 
nature of MF\,83 in \S 4, and summarize our conclusions in \S 5.

\section{Observations}

The datasets used in this study include : (1) deep CCD images taken with 
the Michigan-Dartmouth-MIT (MDM) Hiltner 2.4~m telescope and
the Kitt Peak National Observatory (KPNO) Mayall 4~m telescope,
(2) high-resolution images taken with the {\it Hubble Space Telescope} 
Wide Field/Planetary Camera 2 ({\it HST} WFPC2), 
(3) high-dispersion echelle spectra taken with the KPNO Mayall 4~m telescope, 
and (4) archival {\it R\"ontgen Satellite} ({\it ROSAT}) X-ray images.

\subsection{Ground-based CCD Images}

MF\,83 was observed on 1999 April 12 (UT) with the 
MDM Observatory Hiltner 2.4~m telescope in direct imaging mode.  
The images were recorded with the thinned, backside-illuminated 
$1024\times 1024$ SITE CCD ``Charlotte''.  The pixel size was 24 $\mu$m.
The CCD camera was used at the f/7.5 Cassegrain focus; the field of view 
was 4\farcm7$\times$4\farcm7 and the image scale was 0\farcs275 pixel$^{-1}$.
Images of MF\,83 were obtained in the Johnson {\it B}, {\it V},
and {\it I} filters and
an \ha\ filter ($\lambda_{c}$ = 6563 \AA, $\Delta\lambda$ = 100 \AA).  
The exposure times in {\it B, V, I}, and \ha\ were 1$\times$600~s, 
3$\times$600~s, 2$\times$600~s, and 1$\times$900~s, respectively.  
Standard IRAF\footnote{IRAF
is distributed by the National Optical Astronomy Observatories, which are 
operated by the Association of Universities for Research in Astronomy, Inc.,
under cooperative agreement with the NSF.} tasks were used 
to combine multiple exposures and to remove cosmic rays.  
Cirrus clouds were present during part of the night; therefore, we did not
use these images for photometric measurements.
The seeing was $1\farcs7$, $1\farcs1$, $1\farcs0$, 
and $1\farcs3$ in {\it B, V, I}, and \ha, respectively. 

Deep \ha\ and $R$-band images of MF83 are available from a nova search in 
M101 (Shafter, Ciardullo, \& Pritchet 2000), for which multi-epoch \ha\ 
and $R$-band images were obtained with the KPNO Mayall 4~m telescope in 
1994 May, 1995 April, 1996 May, and 1997 March.  The  2048$\times$2048 T2KB 
CCD detector was used, yielding a 16\arcmin$\times$16\arcmin\ field of view 
and an image scale of 0\farcs47 pixel$^{-1}$.  The \ha\ filter used was 
the KPNO1390, which has a central wavelength of 6567 \AA\ and a 
bandpass of 74 \AA\ FWHM in the converging beam of the telescope. 
The multi-epoch \ha\ images were co-added to yield a super-deep \ha\ image
with a net exposure time of 390 min.  The $R$-band image has an exposure time
of 6 min.

A deep \oiii$\lambda$5007 image of MF83 is available from a survey of 
planetary nebulae in M101 (Feldmeier, Ciardullo, \& Jacoby 1997). The
\oiii\ image was obtained on 1995 April 5 using the same telescope with
the same CCD as the aforementioned deep \ha\ image.  The \oiii\ filter
was centered at 5017 \AA\ with a FWHM of 30 \AA; this central wavelength
was selected to compensate the redshift of M101.
The total exposure time of the \oiii\ image was 120 min.
For continuum subtraction,
an image was also taken with an off-band green continuum filter 
centered at 5312 \AA\ with a FWHM of 267 \AA.  The total
exposure time of this green continuum image is 18 min.

Figure 1 shows a cosmic-ray-cleaned MDM \ha\ image of MF\,83 and its 
environment, and Figure 2 shows MDM images of MF\,83 in the 
{\it B, V, I}, and \ha\ bands.  The \ha\ image in Figure 2d has been 
smoothed with a Gaussian of $\sigma$ = 1 pixel (0\farcs275).  
Figure 3 presents KPNO 4m images of MF\,83 in the \ha\ and \oiii,
red and green continua, and continuum-subtracted \ha\ and \oiii.
Note that these ``\ha'' images are really \ha+\nii\ images, 
as both MDM and KPNO \ha\ filters are broad enough 
to include the \nii$\lambda\lambda$6548, 6583 lines.

\subsection{{\it HST} WFPC2 Images}

The {\it HST} WFPC2 images of MF\,83 were obtained on 1999 June 17 
for the GO program 6928 -- ``The Luminous Giant \hii\ Regions in 
M101.'' As MF\,83 is only $\sim50''$ from the luminous giant \hii\ 
region NGC\,5461, it is serendipitously included in the same Wide 
Field Camera, WF4 in this case.  A Wide Field Camera covers an area 
of 79\farcs7$\times$79\farcs7 and has a pixel size of 0\farcs0996 
pixel$^{-1}$.  As the observations were designed for NGC\,5461, 
which is much brighter than MF\,83, the images are inevitably 
shallow for MF\,83.

The observations were made through the filters {\it F547M} and 
{\it F656N} (see \markcite{Bi96}Biretta et al.\ 1996 for filter 
characteristics) for 2$\times$500 s and 2$\times$600 s, respectively.  
The {\it F547M} filter is centered at 5476.3 \AA\ with a FWHM of 
483.1 \AA.  This filter isolates an emission-line-free band similar 
to, but wider than, the Str\"omgren {\it y} band.  
The {\it F547M} images show stars in and near MF\,83.  
The {\it F656N} filter, centered at 6563.7 \AA\ with a FWHM of 
21.4 \AA, is an \ha\ filter that does not include the \nii\ lines.  
The {\it F656N} images barely detected the \ha\ emission from MF\,83.  
No useful morphological information can be extracted from these 
{\it F656N} images.

The calibrated WFPC2 images were produced by the standard {\it HST} 
pipeline processes.  We processed them further with IRAF and STSDAS 
routines.  The images taken with the same filter were combined to 
remove cosmic rays and to produce a total exposure map.  The combined 
{\it F547M} image was then corrected for the intensity- and 
position-dependent charge transfer efficiency by applying a linear 
ramp with a correction factor chosen according to the average counts 
of the sky background (\markcite{Ho95a}Holtzman et al.\ 1995a).  
The combined {\it F547M} image was further multiplied by the 
geometric correction image f1k1552bu.r9h (downloaded from the 
{\it HST} archives) to compensate the different effective areas 
covered by the pixels (Holtzman et al.\ 1995a; Biretta et al.\ 1996).
This correction is $\sim$ 1-2\% near the edges, with a maximum of 
$\sim$ 4-5\% in the corners of the CCD.

The final corrected $F547M$ image, shown in Figure~4, was used to 
carry out stellar photometry.  We used the APPHOT package in IRAF 
and followed the procedures for photometry detailed by 
\markcite{WH95}Whitmore \& Heyer (1995).  An aperture of 3 pixel
radius (0\farcs3) was used to measure the total counts of each 
stellar source, with the sky brightness determined from a 5-pixel-wide 
annulus of inner radius 10 pixels.  An aperture correction was made
to include all the light within the 0\farcs5 standard aperture, as
recommended by \markcite{Ho95b}Holtzman et al.\ (1995b).
The number of counts ($DN$) in 
each stellar source is converted to an apparent magnitude ($m$) using 
the formula $m = -2.5 \times log_{10}(DN/EXPTIME) + ZEROPOINT$, 
where $EXPTIME$ is the exposure time and 
$ZEROPOINT = -2.5 \times log_{10}(PHOTFLAM) -21.1$.
The parameter $PHOTFLAM$ is $7.747 \times 10^{-18}$
ergs~s$^{-1}$~cm$^{-2}$\AA$^{-1}$ and is a product
provided in the image header by the HST pipeline processes.

The apparent magnitude in {\it F547M}, $m_{F547M}$, can be converted to the
Johnson {\it V} magnitude, $m_V$, by applying a correction of $-$0.023 
mag (appropriate for Vega, \markcite{hst}{\it HST} Data Handbook), 
but this correction is negligible compared to the photometric error 
$\pm$0.2 mag.  Adopting a distance modulus of 29.3 mag (= 7.2 Mpc) 
and a visual extinction $A_V$ = 1.3 mag (Matonick \& Fesen 1997), we 
have converted the {\it V} magnitude to absolute visual magnitude $M_V$.  
The right ascension, declination, $m_{F547M}$, and $M_V$ of the 
sources identified in the {\it F547M} image are listed in Table~2.

\subsection{KPNO Echelle Spectra}

High-dispersion spectra of MF\,83 were obtained with the echelle 
spectrograph on the KPNO Mayall 4 m telescope in two observing runs:
1999 March 2 -- 4 and 1999 June 30 -- July 2 (UT).  The same observing
configuration was used in both runs.
The \mbox{79 line mm$^{-1}$} echelle grating 79-63 and the 
\mbox{226 line mm$^{-1}$} cross disperser 226-1 were used in 
conjunction with the long focus red camera to achieve a reciprocal 
dispersion of \mbox{3.5 \AA\ mm$^{-1}$} at the \ha\ line.
The spectra were imaged with the 2048$\times$2048 T2KB CCD detector.
The 24~$\mu$m pixel size corresponds to 0\farcs24~pixel$^{-1}$ along
the slit and \mbox{$\sim$3.7 \kms\ pixel$^{-1}$} along the dispersion axis.  
Each echellogram covers at least 4,300 -- 7,000 \AA, which includes 
bright nebular emission lines such as \ha, \hb, 
\oiii$\lambda\lambda$4959, 5007, \oi$\lambda$6300,
\nii$\lambda\lambda$6548, 6583, and \sii$\lambda\lambda$6716, 6731.

The journal of observations of MF\,83 is given in Table 1.  
Two orthogonal slit positions, E-W and N-S through the center
of MF\,83, were observed.  Different slit widths were used, 
resulting in different instrumental widths.  The instrumental FWHM, 
determined by Gaussian fits to the unresolved sky lines (telluric OH 
lines), is also given in Table 1.

IRAF software was used for data reduction and analysis.
The reduction processes included bias subtraction, cosmic ray 
removal, and quartz flat-fielding to remove the pixel-to-pixel gain 
variation.  As we were mainly interested in the velocity profiles
and not the surface brightness variation, we did not use a sky
flat to flatten the illumination variation along the slit.
To remove the sky lines, a sky spectrum extracted from regions 
outside MF\,83 is scaled and subtracted from the nebular spectrum.
In order to improve the S/N ratio, the final spectra of MF\,83 
were Hanning smoothed over five adjacent pixels (i.e., weighted 
average with weights of 1/4, 3/4, 1, 3/4, and 1/4 for the 
five pixels).  As the data are oversampled, this smoothing does
not degrade the spectral resolution.

The spectra of MF\,83 obtained in the 1999 March run have low S/N 
ratios, but they show a line split indicating an overall expansion.
The follow-up observations made in the second observing run have 
longer integration times and show the expansion more clearly.  
The new spectra will thus be the main echelle data presented in 
this paper.  Figure 5 displays a portion of the N-S echellogram 
of MF\,83 centered on the \ha\ and \nii$\lambda$6584 lines,
and a close-up of the sky-subtracted \ha\ line.

\subsection{{\it ROSAT} Observations}

Both {\it ROSAT} Position Sensitive Proportional Counter (PSPC) and High
Resolution Imager (HRI) observations are available for MF\,83 (see 
Table~2 of Wang, Immler, \& Pietsch 1999).  These observations are all 
centered at the nucleus of M101. As MF83 is only $\sim$4$'$ from the 
nucleus of M101, the point spread function at MF\,83 should be similar 
to that at the center, $\sim30''$ for the PSPC and $\sim5''$ for the 
HRI ({\it ROSAT} Mission Description 1991).

The HRI observations are merged into a 227 ks exposure.  The merged
HRI image is binned by a factor of two to obtain a pixel size of 1$''$ 
pixel$^{-1}$, and smoothed with a Gaussian of $\sigma$ = 2 pixels (1$''$).  
The smoothed HRI image is presented in Figure 6a, and the X-ray contours 
are overplotted on the KPNO 4~m \ha\ image in Figure 6b.

The PSPC observation has a shorter exposure time, 33.9 ks, and a poorer
angular resolution; therefore, the X-ray source at MF\,83 is not as well 
defined in the PSPC observation as in the HRI image.  The PSPC data have
been integrated over all energy channels (0.1 -- 2.4 keV), binned by a 
factor of 10 to obtain a pixel size of 5$''$ pixel$^{-1}$, and smoothed 
with a Gaussian of $\sigma$ = 2 pixels (10$''$).  The smoothed PSPC 
image is shown in Figure 6c, and the X-ray contours are overplotted on
the KPNO 4~m \ha\ image in Figure 6d.

\section{Analysis \& Results}

\subsection{Interstellar and Stellar Environment of MF\,83}

MF\,83 is located between two spiral arms of M101 (Figure 1). 
Having a \sii/\ha\ ratio of 0.76 and a diameter of $\sim$ 270 pc, 
MF\,83 is one of the three largest SNRs identified in M101 
(Matonick \& Fesen 1997).  It ought to be noted that MF\,83 
is not a known nonthermal radio source; therefore, strictly 
speaking, MF\,83 should be called a SNR candidate.

The ionized gas in MF\,83 is best revealed by the continuum-subtracted
\ha\ image (Figure 3c).  MF\,83 apparently consists of an ionized
gas shell and a bright \hii\ region to the northwest of the shell.
the shell is elongated in the NE-SW direction, with an apparent 
opening at the SW end.  The two bright patches of emission along
the south rim of the shell have stellar counterparts, indicating
that they are \hii\ regions. The continuum-subtracted \oiii\ image
of MF\,83 (Figure 3f) shows a different morphology of the ionized
gas.  The \hii\ regions are much less prominent in the \oiii\ than
in the \ha.  The ionized gas shell of MF\,83, having higher excitation,
is better depicted by the continuum-subtracted \oiii\ image.
We have used \markcite{KG96}Kennicutt \& Garnett's (1996) 
spectrophotometry of \hii\ regions near NGC\,5461 to calibrate the
continuum-subtracted \ha\ and \oiii\ images and made an \oiii/\ha\
ratio map of MF\,83 (Figure 7).  It can be seen that the bright 
\hii\ regions have low \oiii/\ha\ ratios, while regions with low 
surface brightnesses have higher \oiii/\ha\ ratios.

The KPNO 4m green continuum (off-\oiii) image in Figure 3e shows four
sources, a-d, in MF\,83.  Source a is located near the center of the gas 
shell, while sources b-d are in the \hii\ regions along the south rim 
of the shell.  The {\it HST} WFPC2 $F547M$ image of MF\,83 (Figure 4) 
reveals that all of these four sources are extended and must be OB 
associations.  The discrete sources detected in the WFPC2 $F547M$ 
image are marked in Figure 4; their $m_{F547M}$ and $m_{V}$ magnitudes 
are tabulated in Table 2.  The patch of continuum emission at the shell 
center (source a in Figure 3e) is resolved into sources 7, 8, 10, 11, 
and 12 (marked in Figure 4), among which sources 7, 11, and 12 have 
asymmetric and extended images indicating that they are multiples.  
The three concentrations of stellar emission along the southern rim 
of the shell (sources b--d in Figure 3e) are not fully resolved in the 
WFPC2 $F547M$ image, but they (sources 4, 5, and 13) each show irregular,
extended spatial structure indicating the existence of multiple stars.  
Within the compact \hii\ region at the northwest periphery of the shell,
stars are also detected, but they are not as highly concentrated or as 
bright as the stars within the shell and along the south rim of the shell.

The SNR MF\,83 presumably corresponds to the shell structure seen
in the \ha\ image.  The concentration of stars at the shell center 
probably belongs to an OB association.  The three concentrations of 
stars along the shell rim may represent three distinct OB associations, 
as their separations are greater than 70 pc and typical sizes of 
OB associations are $\le$ 70--100 pc (Lucke \& Hodge 1970; Magnier 
et al.\ 1993).  There is an additional OB association in the
compact \hii\ region at the northwest rim of the shell.  
The arrangement of these OB associations and \hii\ regions around
the shell of MF\,83 is reminiscent of the superbubble N11 (Parker 
et al.\ 1992; Mac Low et al.\ 1998) in the Large Magellanic Cloud
(LMC).  MF\,83 is clearly in an environment with star formation.

\subsection{Physical Properties of the Expanding Shell of MF\,83}

The echelle observations of MF\,83 show that the shell structure 
shown in the \ha\ and \oiii\ images is indeed an expanding shell.
MF\,83 is detected in \ha, \nii$\lambda\lambda$6548,6584, and 
\sii$\lambda\lambda$6716,6731 lines.  The spectral region of \ha\ 
and \nii$\lambda$6584 lines is displayed in the upper panel of 
Figure 5, and the sky-subtracted \ha\ line is displayed in 
the lower panel of Figure 5.  These line images show clearly a 
position-velocity ellipse indicating an expanding shell structure.
Using the 2$\sigma$ contour level in the sky-subtracted \ha\
line image, we estimate an expansion velocity of $\sim$ 50 \kms\  
for both the approaching and receding sides of the shell.
We will assume that this expanding shell is associated with the 
supernova remnant in MF\,83, and evaluate its mass and kinetic 
energy.  

The shell mass can be derived from the \ha\ luminosity and
shell thickness.  The \ha\ luminosity of MF\,83, computed 
from the \ha\ flux reported by Matonick \& Fesen (1997), is
1.7$\times 10^{38}$ ergs s$^{-1}$, for a distance of 7.2 Mpc.
However, this \ha\ luminosity includes contributions from the
\hii\ regions along the periphery of the expanding shell; these
\hii\ regions do not participate in the expansion and hence do 
not contribute to the shell kinetic energy.  To exclude the \hii\
regions, we use the \ha\ image in conjunction with the \oiii\ 
image.  As the \hii\ regions have much lower \oiii/\ha\ ratios 
than the shell (see Figure 7), we assume that the \oiii\ flux is 
dominated by the shell emission, and use the \oiii/\ha\ ratio in
the ``clean'' part of the shell to convert the \oiii\ flux to
\ha\ flux.  The \oiii/\ha\ ratio of the entire MF\,83 
is $\sim$0.5, and the \oiii/\ha\ of the clean part of the shell 
is $\sim$0.7.  Using these values and the total \ha\ flux of 
MF\,83, we derive an \ha\ luminosity of 1.2$\times 10^{38}$ ergs 
s$^{-1}$ for the expanding shell.  (Note that this value still 
over-estimates the shell luminosity, as we have assumed that all 
\oiii\ emission originates in the shell.)

The \ha\ image of MF\,83 shows a shell of diameter $\sim$270 pc and 
a fractional shell thickness ($\Delta R/R$) $<$0.5.  
The rms density and mass can be derived from these shell 
parameters and the \ha\ luminosity.  We have made calculations
for $\Delta R/R$ = 0.1 and 0.5, 
the smallest and largest possible shell thicknesses.
We find that the rms density is $\sim$2.0--1.1 H cm$^{-3}$ and the 
shell mass is $\sim$ 1.4--2.5 $\times 10^{5}$ M$_\odot$. 
Using this mass and an expansion velocity of 50 \kms,
we find MF\,83's shell kinetic energy to be
3.5--6.3$\times$10$^{51}$ ergs.  Finally, if the shell of MF\,83
consists of swept-up interstellar medium, the ambient density 
would be 0.6--1.0 H cm$^{-3}$.  These  physical
properties of MF\,83 shell are summarized in Table~3.

\subsection{X-ray Emission and Hot Ionized Gas of MF\,83}

The {\it ROSAT} HRI observations of MF\,83 detected 118$\pm$20 counts 
in the 227-ks exposure.  This number of counts (background-subtracted) is in
perfect agreement with that reported by Wang, Immler, \& Pietsch (1999; 
111$\pm$14 counts for
source H36).  Because of this small number of counts, it is difficult 
to determine accurately whether the X-ray emission is totally unresolved 
or contains both an unresolved component and a slightly extended component.
Figure 6b shows that the position of the X-ray source coincides almost 
perfectly with the optical boundary of MF\,83.

The {\it ROSAT} PSPC observation of MF\,83 detected 64$\pm$14 counts
(background-subtracted) in the 33.9-ks exposure.  This number of 
counts is 25\% higher than that reported by Wang et al.\ (1999),
51$\pm$11 counts,
because we used a low background at $\sim$2$'$~NE of MF\,83 and
our counts may have included more diffuse background within the 
source aperture.  Nevertheless, our number and Wang et al.'s number 
still agree within the error limits.  The observed PSPC count rate 
of MF\,83 is totally consistent with the observed HRI count rate, 
if we adopt the PSPC to HRI count rate ratio of 3 observed in a 
wide range of X-ray sources \markcite{Da96}(David et al.\ 1996).  
The small number 
of PSPC counts prohibits meaningful spectral model fits.  Compared to 
other X-ray sources in M101 (Table 5 of Wang et al.\ 1999), the X-ray 
emission from MF\,83 appears to have a highest hardness ratio HR2
(0.79$\pm$0.22),
defined to be ``(hard1 - hard2)/(hard1 + hard2)'' with ``hard1" 
covering channels 52-90 ($\sim0.75$ keV) and ``hard2" covering 
channels 91--201 ($\sim1.5$ keV).  While this hardness ratio is 
uncertain, it may hint the presence of a hard X-ray component.  
Future observations with a deeper exposure, a higher angular 
resolution, and a spectral range covering higher photon energies 
are needed to resolve the X-ray sources at MF\,83.

The PSPC image suggests that the discrete source in MF\,83
detected by the HRI is superposed on a diffuse background.
As we do not have an effective way to separate the emission of MF\,83
from the diffuse background emission, we will use the HRI count rate
to represent the emission from MF\,83.  We assume that all X-ray emission
from MF\,83 originates from a plasma at 3$\times$10$^6$ K.  (This is 
a reasonable assumption because hot interstellar gas is most frequently
observed to be a few 10$^6$ K, and the X-ray emissivity varies very
slowly with temperature at this temperature range.)
Using a reddening of $E(B-V)$ = 0.42 mag for MF\,83 (Matonick \& Fesen 
1997) and a Galactic gas to dust ratio of $N_{\rm H}/E(B-V)$ =
5.8$\times10^{21}$ H cm$^{-2}$ mag$^{-1}$ (Bohlin, Savage, \& Drake 
1978), we obtain a hydrogen column density of 2.4$\times$10$^{21}$ 
H cm$^{-2}$.  Adopting this hydrogen column density as an approximation 
for the absorption column density and a 3$\times$10$^6$ K plasma temperature, 
we find the HRI count rate corresponding to an unabsorbed X-ray
flux of 1.1$\times$10$^{-13}$ erg cm$^{-2}$ s$^{-1}$ in the 0.1--2.4
keV band\footnote{calculated by using W3PIMMS tool at 
http://heasarc.gsfc.nasa.gov/Tools/w3pimms.html}.  
For a distance of 7.2 Mpc to M101, the X-ray luminosity 
of the X-ray source in MF\,83 is 6.8$\times$10$^{38}$ erg s$^{-1}$.
This luminosity is about five times as high as that reported by Wang 
(1999), 1.2$\times$10$^{38}$ erg s$^{-1}$, 
because Wang et al.\ assumed a smaller absorption column
density and used a smaller counts-to-energy conversion factor.

Our X-ray flux of MF\,83 implies an emissivity of $\sim$2.5$\times$10$^{-23}$ 
erg cm$^3$ s$^{-1}$ for a 3$\times$10$^6$ K plasma with solar abundances.
If we assume that the X-ray emitting gas is concentrated in a shell with a
certain fractional shell thicknesses ($\Delta R/R$), we may derive the rms 
density, mass, and thermal energy of the X-ray emitting gas.   If 
$\Delta R/R$ = 0.1, the rms density, mass, and thermal energy will
be 0.6 H cm$^{-3}$, 4$\times$10$^{4}$ M$_\odot$, and 2.9$\times$10$^{52}$ 
ergs, respectively; if $\Delta R/R$ = 0.5, these figures will be
0.3 H cm$^{-3}$, 7$\times$10$^4$ M$_\odot$, and 5.2$\times$10$^{52}$ ergs, 
respectively.

\section{Discussion}

The SNR MF\,83 in M101 is intriguing because it has been suggested to be 
a hypernova remnant and it may need an explosion energy comparable to 
those associated with GRBs (Wang 1999).  Our new observations 
allow us to examine critically the physical nature of MF\,83.  We will 
first compare MF\,83 to known, common astronomical objects and assess
the likelihood that MF\,83 has a similar nature.  We will then consider
the possibility of the existence of an energetic SNR and evaluate
its energetics.

\subsection{MF\,83 as a Star Formation Region}

MF\,83 consists of a $\sim$270 pc ionized gas shell, three \hii\ regions 
along the south rim and another \hii\ region outside the northwest rim
of the shell.  This arrangement is commonly seen in star formation
regions.  These \hii\ regions individually have an \ha\ luminosity of 
$\sim$2$\times$10$^{37}$ ergs~s$^{-1}$.  This \ha\ luminosity requires 
an ionizing flux of $\sim$2$\times$10$^{49}$ ionizing photons s$^{-1}$, 
which can be easily supplied by an O6 star or several later-type O stars.  
The presence of late-type O stars, or a lack of early-type O stars, 
is also supported by the low \oiii/\ha\ ratios of the \hii\ regions.  

The stellar continuum sources detected in the WFPC2 $F547M$ images 
have absolute visual magnitudes $M_V$ in the range of $-6$ to $-8$ 
mag (Table~2).
While the faintest sources might be single early-type supergiants,
the brighter sources and especially the sources showing asymmetric 
or extended images are composites, e.g., sources 4, 5, 12, and
13 in Table~2 and Figure 4.  The composite sources embedded in \hii\ 
regions (sources 4, 5, and 13) and the composite source at the shell 
center (source 12) are most likely OB associations.

The \hii\ regions and OB associations in MF\,83 are only about
twice as powerful as the Orion Nebula, and are comparable
to the less impressive OB/\hii\ regions in the LMC.  MF\,83
is a star formation region, but by no means a starburst region.

\subsection{The Supernova Remnant Candidate in MF\,83}

The existence of a SNR in MF\,83 is suggested by the high \sii/\ha\ 
ratio, 0.76 (Matonick \& Fesen 1997), and bright X-ray emission, 
1.2$\times$10$^{38}$ ergs s$^{-1}$ (Wang 1999).  These two properties, 
unfortunately, cannot unambiguously ensure the existence
of a SNR for the following reasons.  First, the \sii/\ha\ ratio is 
high for both low-velocity shocks (shock velocity $\le$50 \kms) and 
high-velocity shocks (shock velocity $\ge$100 \kms) 
\markcite{SM79}(Shull \& McKee 1979).  While SNRs with large shock 
velocities show high \sii/\ha\ ratios, superbubbles with expansion 
velocities approaching 50 \kms\ also show high \sii/\ha\ ratios,
for example, the superbubbles N185 and N186E in the LMC
have \sii/\ha\ ratios of 0.64 and $\sim$1, respectively
\markcite{La77}(Lasker 1977).  
Second, the X-ray source in MF\,83 is not resolved by the $ROSAT$ 
HRI or PSPC.  Moreover, the PSPC spectral hardness ratio of MF\,83 
suggests the existence of a hard component, which is frequently 
associated with X-ray binaries, and the X-ray luminosity of MF\,83 
is within the range of luminosities seen in X-ray binaries in M101 
(Wang et al.\ 1999).  Therefore, until the X-ray source is 
shown to be spatially extended or the radio spectral index is 
observed to be nonthermal, the existence of a SNR in MF\,83 is not 
confirmed.

\subsection{The Large Expanding Shell as an X-ray-bright Superbubble}

The large shell in MF\,83 was initially identified as the SNR (Matonick 
\& Fesen 1997).  As we have illustrated in \S4.1, this large shell is 
centered on groups of stars that are likely to be OB associations.  
The large shell size and the central stellar content suggest that this
shell is a superbubble.  
It is then interesting to compare the physical properties of this
large shell in MF\,83 to those of known superbubbles in the LMC,
which have been extensively studied.  Most LMC superbubbles with sizes
greater than 100 pc have expansion velocities well below 50 \kms\
\markcite{Ge83}(Georgelin et al.\ 1983; \markcite{Ro86}Rosado 1986),
with N185 being the only exception (diameter $\sim$ 105 pc, expansion
velocity = 70$\pm$10 \kms, \markcite{Ro82}Rosado et al.\ 1982).
The 50 \kms\ expansion velocity and 270-pc diameter of MF\,83 thus
make it unusual: it is unusually large for its expansion velocity,
or it expands unusually fast for its size.

The expansion velocity of the MF\,83 shell implies a shock velocity 
of $\sim$67 \kms.  The post-shock region should have a high \oiii/\ha\ 
ratio, but not larger than 1.0 \markcite{SM79}(Shull \& McKee 1979; 
\markcite{Ha87}Hartigan, Raymond, 
\& Hartmann 1987).  This is consistent with the observed high \oiii/\ha\
ratio in the shell, $\sim$0.7.  This shock velocity may also be responsible 
for the high \sii/\ha\ ratio observed in the MF\,83 shell, but is not 
high enough to produce X-ray emission in the post-shock region.  
The high \sii/\ha\ ratio and bright X-ray emission must not originate from 
the same shocks.  If the X-ray emission from MF\,83 is indeed diffuse 
and originates in the shell interior, the MF\,83 shell would be similar 
to the X-ray-bright superbubbles in the LMC, which have been suggested to 
be powered by interior SNRs near the shell walls \markcite{CM90}(Chu 
\& Mac Low 1990; \markcite{WH91}Wang \& Helfand 1991).  
Many X-ray-bright superbubbles in the LMC are known to expand faster 
than expected and have high \sii/\ha\ ratio \markcite{Oe96}(Oey 1996).
It is possible that the large shell in MF\,83 is an X-ray-bright 
superbubble.

As an X-ray-bright superbubble, the MF\,83 shell is unusually large
for its expansion velocity and is exceptionally luminous in X-rays.
Compared to the X-ray-brightest superbubble in the LMC, N44 
\markcite{Ch93}(Chu et al.\ 1993), the MF\,83 superbubble is $\sim$500 
times more luminous.  The internal thermal energy implied by the 
X-ray emission is a few $\times$10$^{52}$ ergs.  It is conceivable that
the high thermal pressure, resultant from this large thermal energy,
drives the fast expansion of the superbubble.  However, this amount of
thermal energy is more than an order of magnitude higher than the
canonical explosion energy of a supernova.   If the large shell in 
MF\,83 is an X-ray-bright superbubble, it requires either a 
``hypernova" or 10--100 normal supernovae within the past 
$\sim$ 1$\times$10$^6$ years, the cooling time scale for a hot gas 
with a temperature of 3$\times$10$^6$ K and density of 0.5 H cm$^{-3}$.

\subsection{The Large Expanding Shell as a Hypernova Remnant}

Finally, we consider the possibility that the large shell of MF\,83 
is formed by one single energetic explosion.  Our analyses in 
\S3.2 and \S3.3 show a kinetic energy of a few $\times$10$^{51}$ ergs
in the ionized gas shell, and an internal thermal energy of a few 
$\times$10$^{52}$ ergs in the shell interior.  A similar disparity
between the ionized shell kinetic energy and internal thermal
energy has been seen in the X-ray-bright superbubble N44, but it is 
found that the ionized shell of N44 is enveloped by a co-expanding,
more massive, neutral \hi\ shell \markcite{Ki98}(Kim et al.\ 1998).  
It is possible that the MF\,83 shell also has a neutral shell carrying
more mass and kinetic energy than the ionized shell.  The kinetic
energy in the ionized gas shell is thus a lower limit on the shell
kinetic energy, and cannot be used to determine the explosion 
energy.

If the MF\,83 shell is formed by one single explosion, the 
explosion energy must be greater than the observed internal thermal
energy, and may approach $10^{53}$ ergs, which is about two orders 
of magnitude higher than the canonical explosion energy for a normal
supernova (Jones et al.\ 1998).  This large energy requirement
would qualify MF\,83 as a ``hypernova remnant."  However, to prove
that MF\,83 is indeed a hypernova remnant, it is necessary to 
demonstrate convincingly that a single explosion formed the MF\,83 shell.
The existence of OB associations within MF\,83 provides circumstantial
evidence that multiple explosions produced the shell, but this neither
demonstrates or rules out that a dominant powerful explosion occurred.

\section{Summary and Conclusions}

The SNR candidate MF\,83 in M101 has been suggested to be a 
hypernova remnant that requires an explosion energy about two
orders of magnitude higher than those of normal supernovae,
based mainly on the high X-ray luminosity of MF\,83 and the 
assumption that all X-ray emission originates from shock-heated
gas in the remnant.  

We have analyzed high-quality ground-based and $HST$ images of 
MF\,83, and found that MF\,83 is a star formation region.  The
\ha\ images show that MF\,83 consists of four \hii\ regions and
a large ionized shell.  The stellar continuum images show groups
of stars in the \hii\ regions and within the shell; these groups 
of stars are most likely OB associations.

The stellar content and physical properties of the large shell
in MF\,83 suggest that it is a superbubble.  If the X-ray emission
is indeed diffuse, this shell will be an X-ray-bright superbubble
whose interior has been heated by recent supernovae.  The thermal
energy derived from the X-ray data is a few $\times$10$^{52}$ ergs,
which requires either a ``hypernova" or 10--100 normal supernovae 
in the past 10$^6$ years.  If it can be proved that the shell of 
MF\,83 is produced by one single explosion, then a ``hypernova" 
may be in order.

Future X-ray observations with high angular resolution is needed
to resolve the X-ray source in MF\,83.  Only after excluding
point sources in MF\,83 can we determine the true amount of 
diffuse X-ray emission, re-evaluate the energy budgets, and 
assess the ``hypernova remnant" nature of MF\,83.

\acknowledgments

This research is partially supported by the grant STScI GO-6829.01-95A, 
and uses observations with the NASA/ESA {\em Hubble Space Telescope}, 
obtained at the Space Telescope Science Institute, which is operated by 
the AURA, Inc., under NASA contract NAS5-26555.  This paper also uses
data obtained (in part) with the 2.4 m Hiltner Telescope of the 
Michigan-Dartmouth-MIT Observatory.
SPL thanks the support by the Laboratory for Astronomical Imaging
through NSF grants AST 96-13999 and AST 98-20641.
EKG gratefully acknowledges support by NASA through grant HF-01108.01-98A 
from the Space Telescope Science Institute.

\newpage

\begin{table}[th]
\caption{Journal of KPNO 4m Echelle Observations}
\vskip 0.5cm
\begin{tabular}{ccccc}
\hline \hline
Date of observation & Slit orientation & Slit width&  Exposure time& Instrumental FWHM \\
\hline
1999 March 4 & NS & 2\farcs0 &  1$\times$1800~s & 19.0$\pm$1.6~\kms \\
1999 June 30 & EW & 1\farcs5 &  4$\times$1800~s & 14.9$\pm$0.6~\kms\\
1999 July 1  & NS & 2\farcs0 &  3$\times$1800~s & 18.3$\pm$0.5~\kms\\
\hline\hline
\end{tabular}
\end{table}

\begin{center}
\begin{table}[p]
\caption{Photometry of Bright Stellar Sources in MF83}
\vskip 0.5cm
\begin{tabular}{rcccc}
\tableline
\tableline
ID      & R.A.    & Decl.   & $m_{F547M}$\tablenotemark{a} & $M_{V}$\tablenotemark{b}  \\
        & (J2000) & (J2000) &    mag                       &     mag            \\
\tableline

1   &14  3 35.67 &54 19 26.92 & 24.6&  -6.1\\
2   &14  3 35.76 &54 19 29.27 &  24.3  &  -6.3 \\
3   &14  3 35.77 &54 19 18.85 &  24.9  &  -5.8 \\
4   &14  3 35.80 &54 19 20.90 &  22.6  &  -8.1 \\
5   &14  3 35.98 &54 19 19.88 &  22.9  &  -7.8 \\
6   &14  3 36.05 &54 19 20.76 &  24.7  &  -6.0 \\
7   &14  3 36.05 &54 19 22.78 &  23.4  &  -7.3 \\
8   &14  3 36.05 &54 19 24.34 &  24.5  &  -6.1 \\
9   &14  3 36.09 &54 19 19.46 &  24.7  &  -6.0 \\
10  &14  3 36.15 &54 19 25.71 &  24.8  &  -5.9 \\
11  &14  3 36.16 &54 19 24.35 &  24.0  &  -6.7 \\
12  &14  3 36.17 &54 19 23.41 &  22.7  &  -7.9 \\
13  &14  3 36.35 &54 19 20.46 &  23.2  &  -7.4 \\
14  &14  3 36.66 &54 19 23.15 &  23.4  &  -7.2 \\
15  &14  3 36.73 &54 19 22.25 &  24.3  &  -6.4 \\

\tableline
\tableline

\tablenotetext{a}{Apparent magnitude in the WFPC2 $F547M$ bandpass. The photometric error is $\sim$ 0.2 mag.} 
\tablenotetext{b}{Assuming an extinction of $A_V$ = 1.3 mag (Matonick \& Fesen 1997) and a distance of 7.2 Mpc (Stetson et al.\ 1998).}
\end{tabular}
\end{table}
\end{center}

\newpage

\begin{table}
\caption{Physical Properties of MF\,83}
\vskip 0.5cm
\begin{tabular}{lcc}
\hline \hline
Shell thickness ($\Delta R/R$)            & 0.1 & 0.5 \\
\hline
Electron density (cm$^{-3}$)            & 2.0 & 1.1 \\
Mass of the shell ($10^{5}$ M$_\odot$)  & 1.4 & 2.5 \\
Ambient density (cm$^{-3}$)             & 0.6 & 1.0 \\
Kinetic energy ($10^{51}$ ergs)         & 3.5 & 6.3 \\
\hline\hline
\end{tabular}
\end{table}

\clearpage

\begin{figure}
\centerline{\plotone{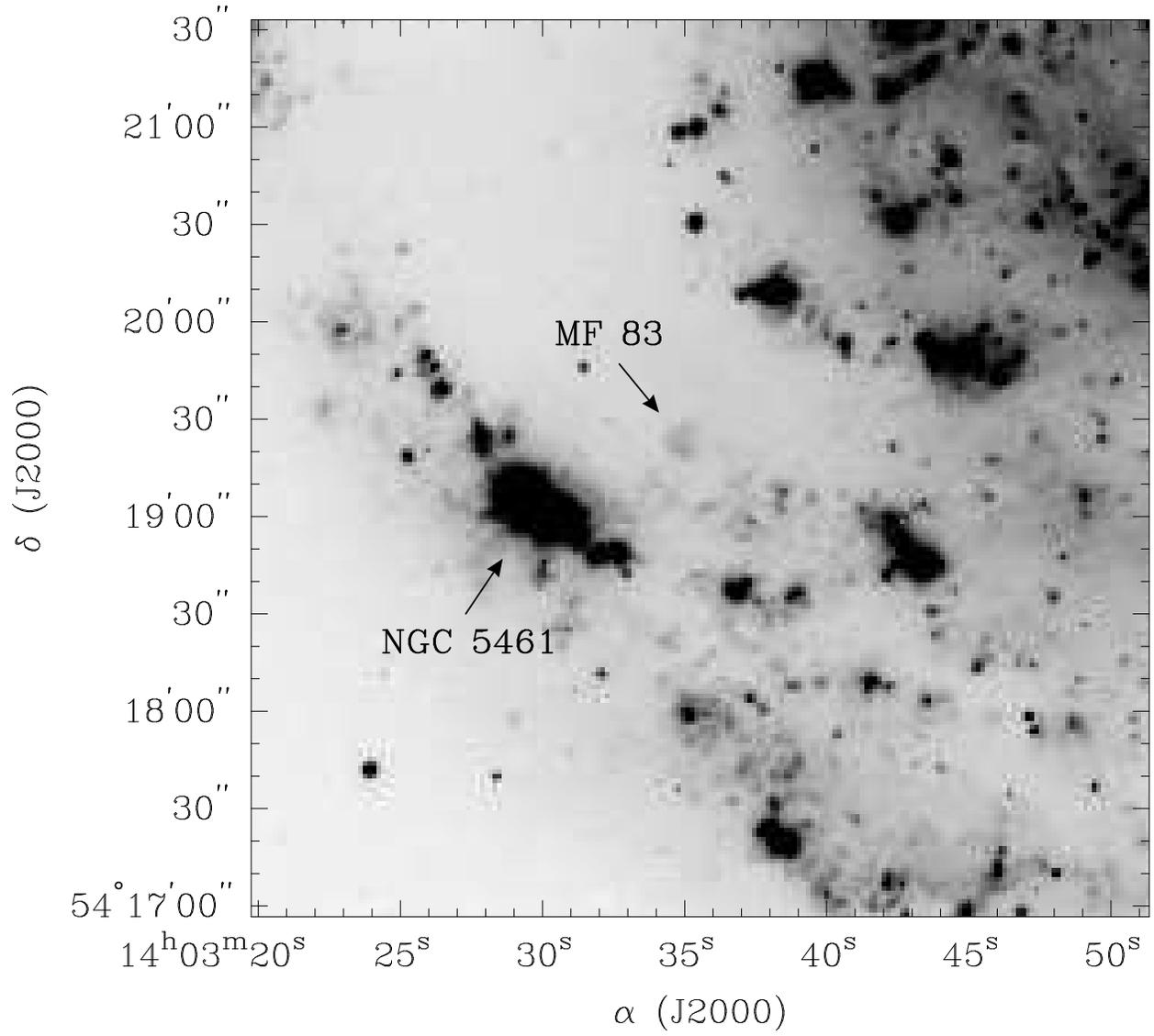}}
\figcaption{MDM 2.4m \ha\ image of MF83 and an eastern portion of M101.}
\end{figure}

\begin{figure}
\plotone{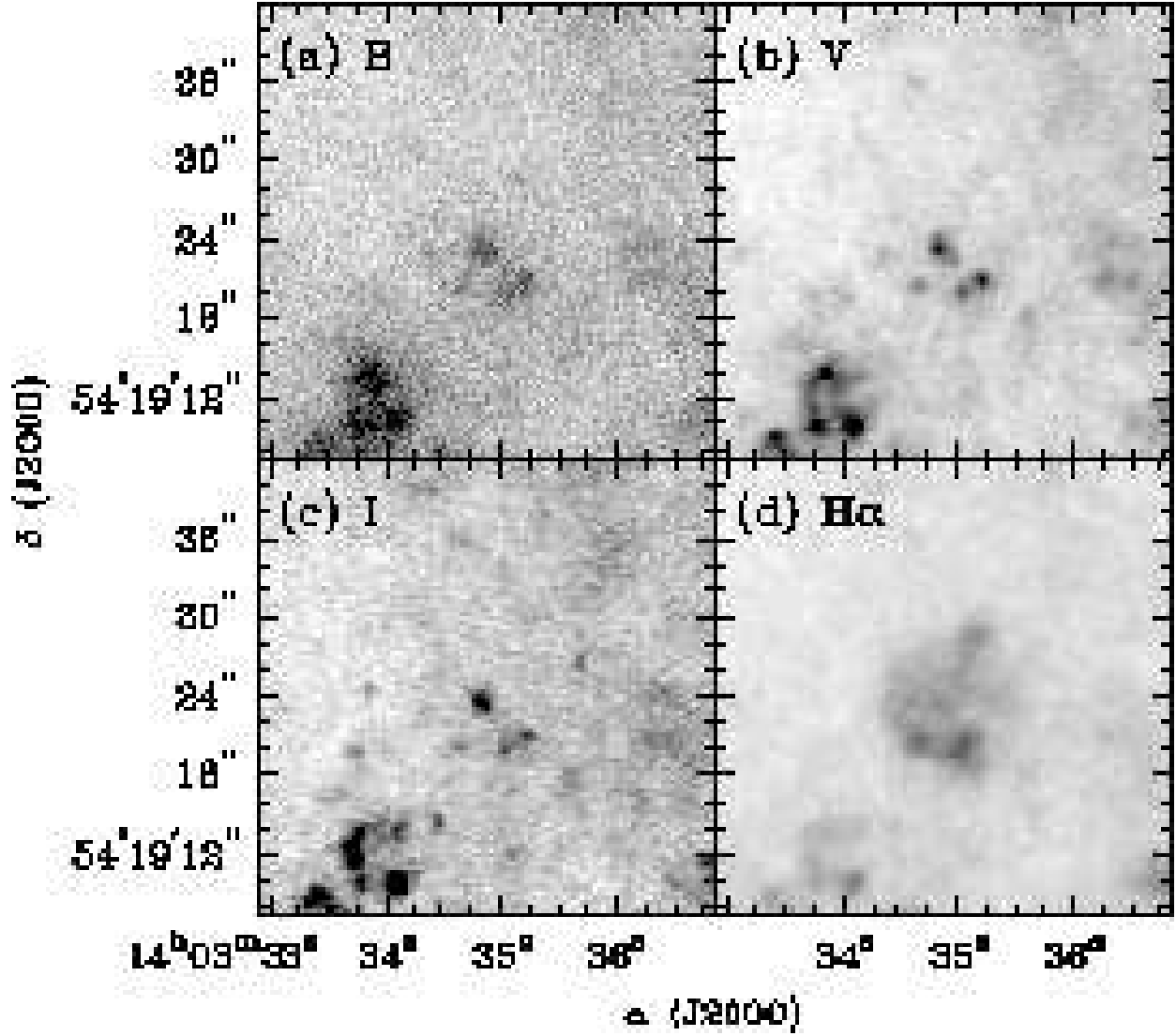}
\caption{MDM 2.4m images of MF\,83 in the (a) {\it B}, (b) {\it V}, (c) {\it I} bands
and the (d) \ha\ line.  The field of view of each plot is $35''\times35''$.}
\end{figure}

\begin{figure}
\epsscale{0.8}
\plotone{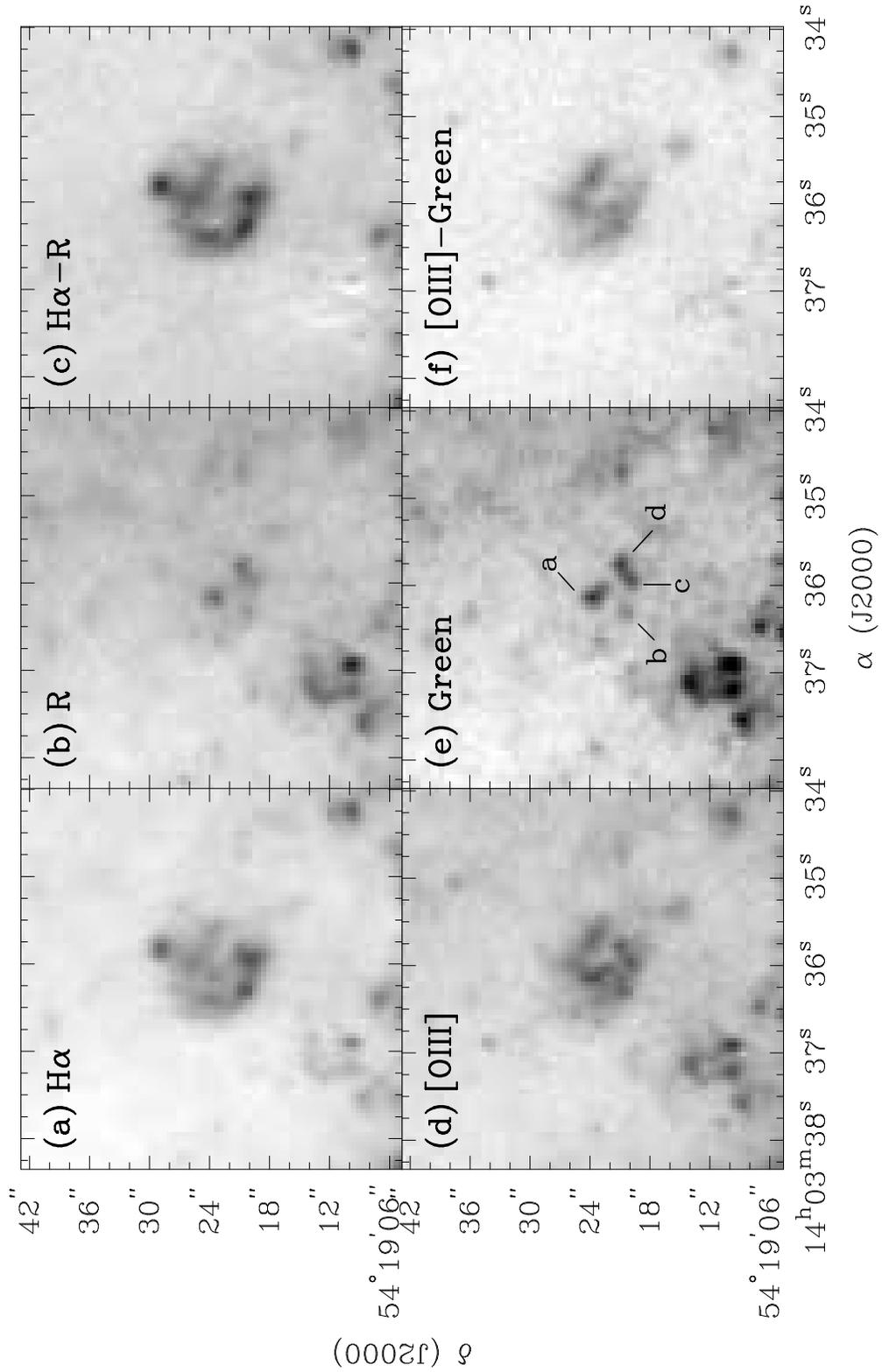}
\caption{KPNO 4m images of MF\,83 in (a) \ha, (b) R band, 
(c) continuum-subtracted \ha, (d) \oiii, (e) green continuum,
and (f) continuum-subtracted \oiii.  Four concentrations of
stars, a--d, are marked in (e).}
\end{figure}

\begin{figure}
\epsscale{0.8}
\plotone{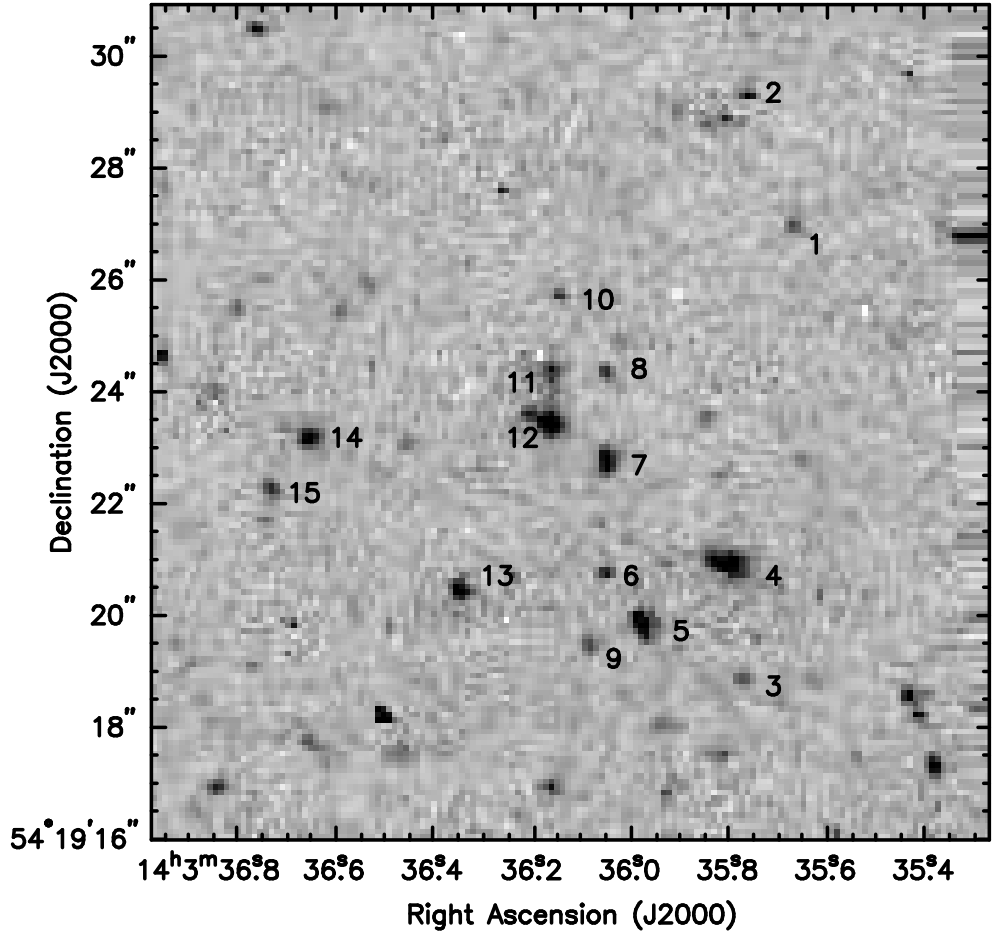}
\caption{$HST$ WFPC2 image of MF\,83 in the $F547M$ band.  The photometry of
 the sources marked is given in Table~2.}
\epsscale{1}
\end{figure}

\begin{figure}
\plotone{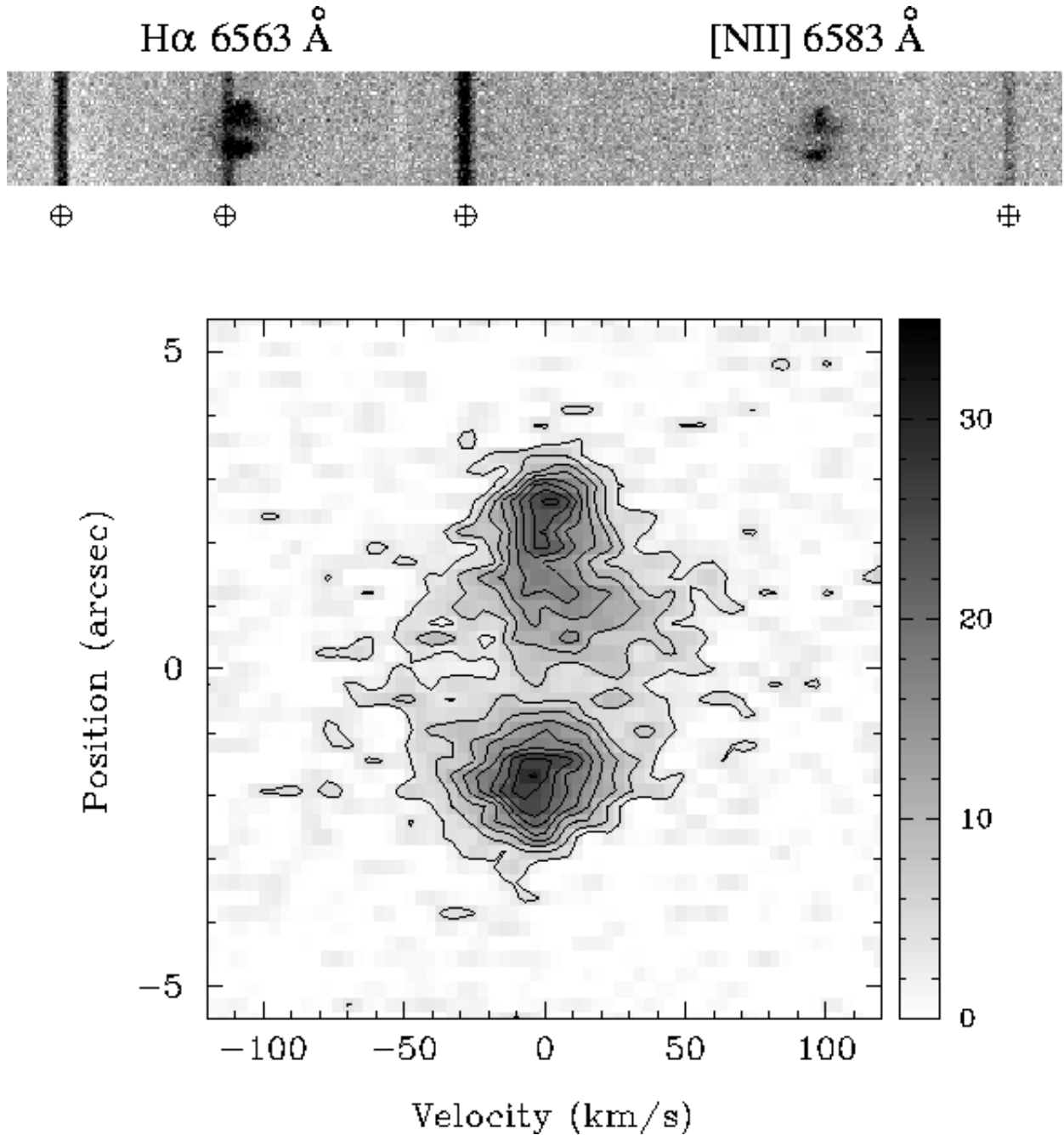}
\caption{Top: KPNO 4m echelle image of the \ha\ and \nii\ lines for a N-S slit
   position centered on MF\,83.  The telluric lines marked by $\oplus$
   are OH 6553.6 \AA, \ha\ 6562.8 \AA, OH 6568.8 \AA, and OH 6577.2/6577.4 \AA.
   Bottom: The sky-subtracted \ha\ line from the above echellogram, Hanning
   smoothed over five pixels along the velocity axis.
   The 1 $\sigma$ noise level is 1.6 counts pixel$^{-1}$.  
   The contours are at 2, 4, 6, 8, 10, 12, 14, and 16 $\sigma$ levels.}
\end{figure}

\begin{figure}
\plotone{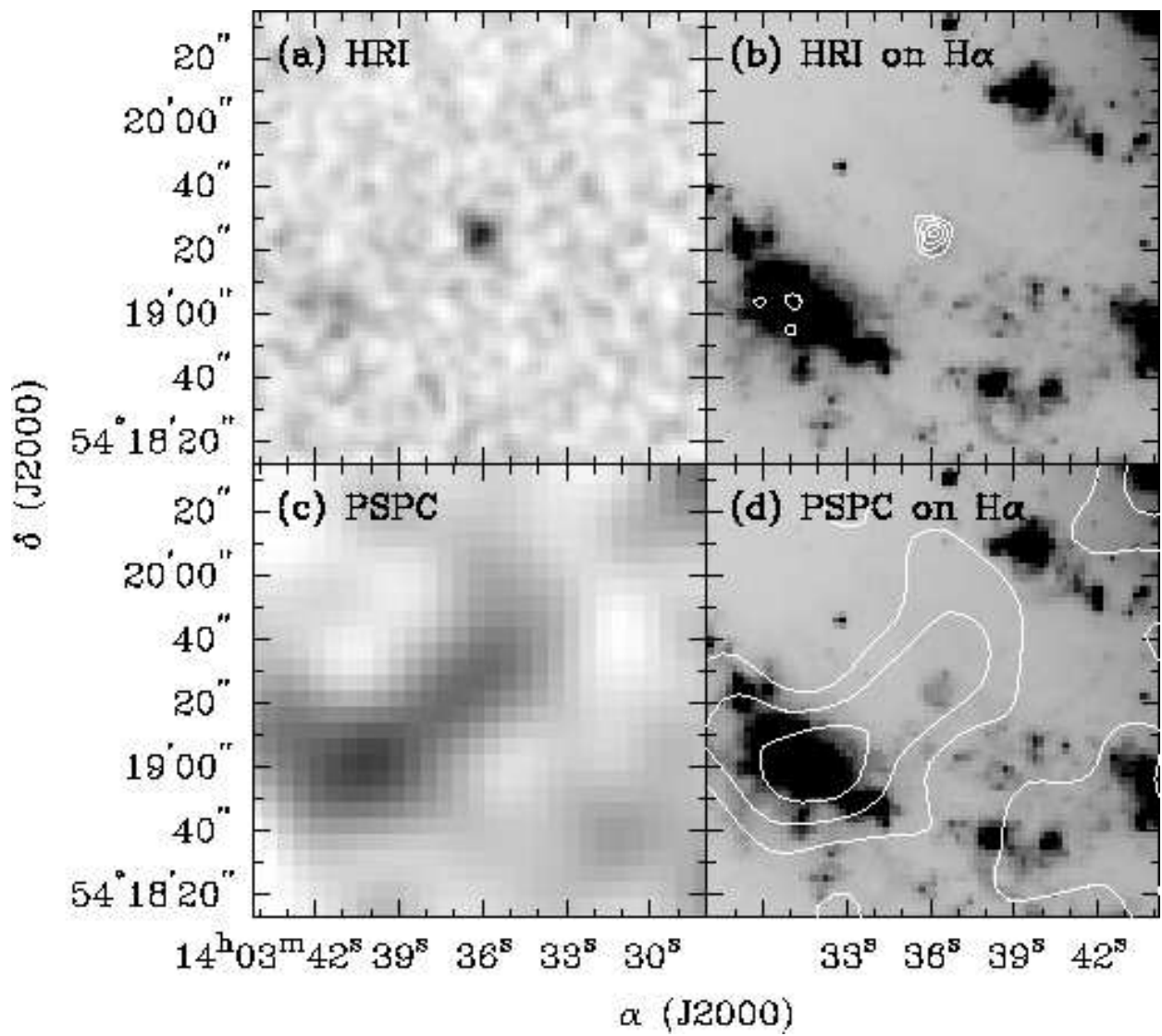}
\caption{$ROSAT$ X-ray images of MF\,83.  (a) HRI image in greyscale. 
This HRI image has been smoothed with a Gaussian of $\sigma$ = 1$''$.
(b) HRI contours over the KPNO 4m \ha\ image.  The contour levels are 45, 
60, 75, 90\% of the peak. (c) PSPC image in greyscale.  This PSPC image 
has been smoothed with a Gaussian of $\sigma$ = 10$''$.
(d) PSPC contours over the KPNO 4m \ha\ image.  The contour levels
are 1.2, 1.5, 1.8 counts pixel$^{-1}$.}

\end{figure}

\begin{figure}
\plotone{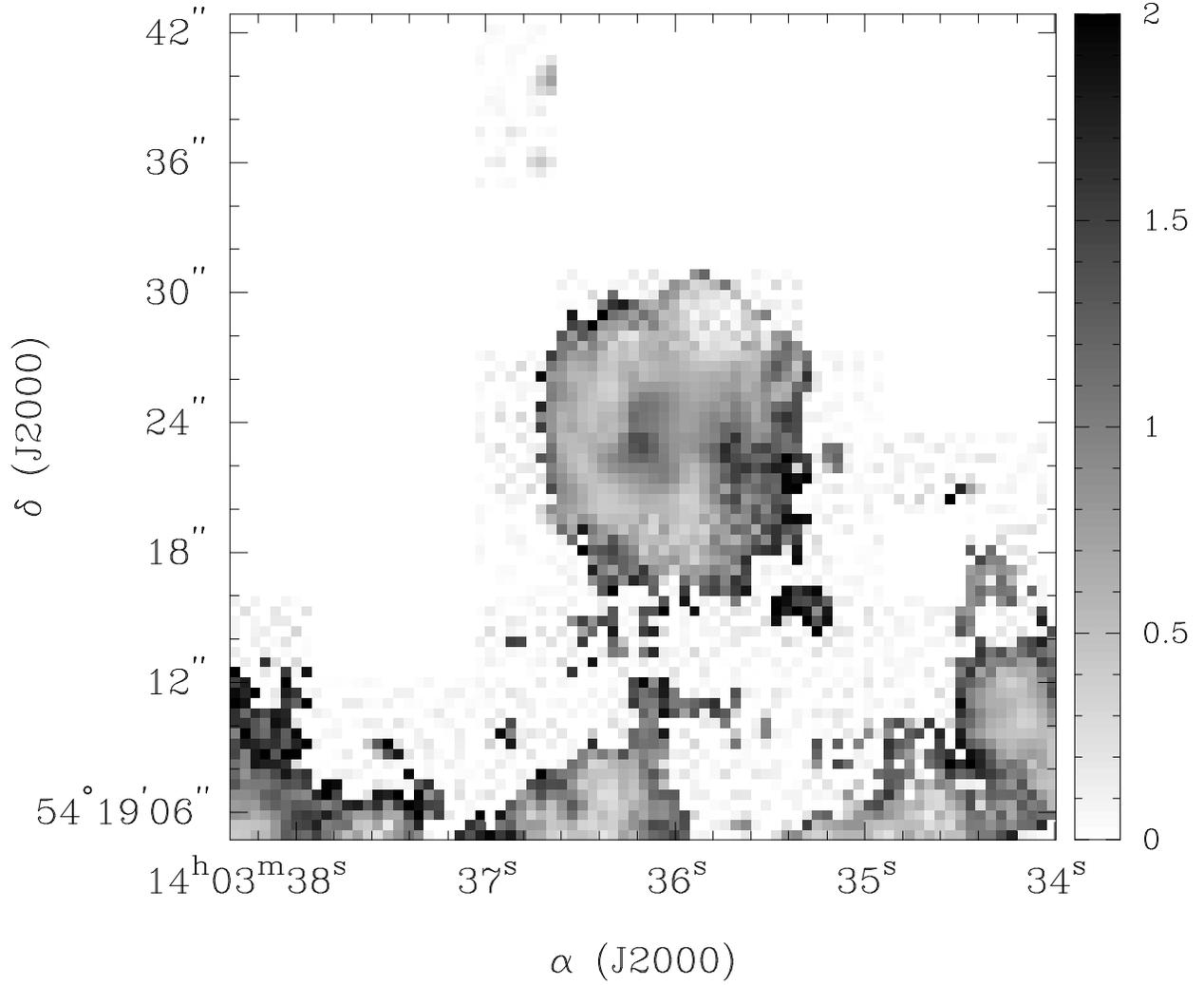}
\caption{\oiii/\ha\ ratio map of MF\,83.  The grey scale bar on the right
shows the \oiii/\ha\ range of 0--2.0.}
\end{figure}

\end{document}